\documentclass[11pt]{article}
\usepackage{amsmath}
\usepackage{amssymb}
\usepackage{amsthm,amsxtra}
\usepackage{epsf}
\usepackage{eepic}
\newlength{\mytopmargin}
\newlength{\myleftmargin}
\newtheorem{lemma}{Lemma}
\newtheorem{prop}[lemma]{Proposition}
\newtheorem{thm}{Theorem}
\newtheorem{cor}[thm]{Corollary}

\begin{document}

\title{\Large\bf Skew orthogonal polynomials and the partly symmetric real Ginibre ensemble} 
\author{Peter J. Forrester$^{\dagger}$ and Taro Nagao${}^*$}
\date{}
\maketitle

\begin{center}
\it
 $^{\dagger}$Department of Mathematics and Statistics, University of Melbourne, \\
Victoria 3010, Australia
\end{center}
\begin{center}
${}^*$
\it
Graduate School of Mathematics, Nagoya University, \\
Chikusa-ku, Nagoya 464-8602, Japan
\end{center} 

\bigskip
\begin{center}
\bf Abstract 
\end{center}
\par
\bigskip
\noindent 
The partly symmetric real Ginibre ensemble consists of  matrices formed
as linear combinations of  real symmetric and real anti-symmetric Gaussian
random matrices. Such matrices typically have both real and complex eigenvalues.
For a fixed number of real eigenvalues, an earlier work has given the explicit
form of the joint eigenvalue probability density function. We use this to derive
a Pfaffian formula for the corresponding  summed up generalized partition function.
This Pfaffian formula allows the probability that there are exactly $k$ eigenvalues
to be written as a determinant with explicit entries. It can be used too to give
the explicit form of the correlation functions, provided certain skew orthogonal polynomials
are computed. This task is accomplished in terms of Hermite polynomials, and allows
us to proceed to analyze various scaling limits of the correlations, including
that in which the matrices are only weakly non-symmetric.

\newpage
\section{Introduction}
In random matrix theory the Ginibre ensembles \cite{Gi65} refer to Gaussian random
matrices with either real, complex, or real quaternion entries, which are
all independent. In distinction to ensembles of
Hermitian matrices, the support of the eigenvalues in the Ginibre ensembles is a disk in
the complex plane. Thus the eigenvalue distribution can be regarded as specifying a
point process in a two-dimensional domain. In this paper we will study the point process
associated with the eigenvalue distribution for random matrices which interpolate between
the real Ginibre ensemble and the Gaussian orthogonal ensemble (GOE) of real symmetric matrices.

The two-dimensional point process associated with the eigenvalue distribution for random matrices
which interpolate between the complex Ginibre ensemble and the 
Gaussian unitary ensemble (GUE) of
complex Hermitian matrices has been the subject of an earlier study \cite{FKS97}.
Similarly, the eigenvalue distribution for
random matrices interpolating between the real quaternion Ginibre
ensemble and the Gaussian symplectic ensemble (GSE) of real quaternion Hermitian matrices
has also been analyzed as a point process \cite{Ka98}.
The interpolating ensemble to be studied herein is thus the last of those naturally associated
with the Ginibre ensembles to be considered from this viewpoint. Such studies are
well motivated for their relevance to Efetov's theory of directional quantum chaos
\cite{Ef97}, in which a special role is played by the interpolating ensembles
in the weak non-Hermiticity limit.

The point processes associated with the Ginibre ensembles have the feature
of being proportional to $e^{-\beta U}$ for potentials $U$ which are the sum
of one and two body terms, the two body terms being logarithmic. As the
logarithmic pair potential is that for two-dimensional charges, there is thus
an analogy with the equilibrium statistical mechanics of certain two-dimensional 
one-component Coulomb systems (see e.g.~\cite{Fo98a} and references therein). This is most
immediate in the case of the complex Ginibre ensemble, for which the eigenvalue
probability density function (PDF) is proportional to $e^{-\beta U}$ with
$\beta = 2$ and
\begin{equation}\label{2.1}
U = {1 \over 2} \sum_{l=1}^N |z_l|^2 - \sum_{1 \le j < k \le N}
\log |z_k - z_j|, \qquad
z_j := x_j + i y_j. 
\end{equation}
The potential (\ref{2.1}) is due to $N$ unit two-dimensional charges, repelling via
the logarithmic pair potential $- \log |z - z'|$, and with a smeared out disk of uniform
neutralizing charge centred about the origin of charge density $-1/\pi$, 
which creates the one-body harmonic
potential ${1 \over 2} |z|^2$. Note that the disk must have radius $\sqrt{N}$ to neutralize
the $N$ mobile charges. As we expect equilibrium Coulomb systems to be locally charge
neutral (otherwise an electric field would be created, and the system would go out of
equilibrium), the particle density should to leading order also be a disk centred about
the origin of radius $\sqrt{N}$, a fact which can be checked upon exact calculation of
the one-point correlation. We remark that recently variants on the eigenvalue problem
for complex Ginibre matrices have been formulated \cite{Kr07}, which have analogies
with one-component Coulomb systems on the surface of a sphere \cite{Ca81}
and in a hyperbolic disk \cite{JT98} (in relation to the latter see too the
work \cite{PV03} on certain random complex polynomials).

The eigenvalue PDF for matrices interpolating between the
complex Ginibre ensemble and the Gaussian unitary ensemble (GUE) also has
a Coulomb gas analogy. Such matrices can be written in the form
$H + i v A$ where $H$ and $A$ are Hermitian matrices from the Gaussian
unitary ensemble (GUE), scaled so that the joint PDF of the
elements is proportional to $ \exp ( - {1 \over 1 + \tau} {\rm Tr} \, X^2 )$,
$\tau = (1 - v^2)/(1+v^2)$. The eigenvalue PDF can
be computed as being proportional to \cite{FKS97}
\begin{equation}\label{11.ell}
\exp \Big ( - {1 \over 1 - \tau^2}
\sum_{j=1}^N \Big ( |z_j|^2 - {\tau \over 2}
(z_j^2 + \bar{z}_j^{2} )\Big ) \Big ) \prod_{1 \le j < k \le N}
| z_k - z_j|^2.
\end{equation}
Here the one-body potential can be interpreted as being due to a uniformly
charged ellipse, charge density $-1/\pi (1 - \tau^2)$, semi-axes $A$ and $B$ given
by  \cite{DGIL94,FJ96}
\begin{equation}\label{AB}
A = \sqrt{N} (1 + \tau), \qquad B = \sqrt{N} (1 - \tau).
\end{equation}
 Again, one can anticipate
that the particle density will to leading order be of this same shape, a fact which
can be verified by exact computation of the one-point correlation \cite{DGIL94}. 

As for the complex Ginibre ensemble, the leading eigenvalue support of the real Ginibre
ensemble is also a disk \cite{Ed95}. Moreover, in the case of matrices interpolating
between the real Ginibre ensemble and the 
GOE, it has
been anticipated that the support will be an ellipse \cite{LS91}. By exact calculation
of the one-point function, verification of this fact is given in Section 5.1 below.

We begin in Section 2 by defining matrices interpolating between the real Ginibre
ensemble and the GOE in terms of a linear
combination of random real symmetric and anti-symmetric matrices. Knowledge of
the eigenvalue PDF for the real Ginibre ensemble allows the eigenvalue PDF of
such matrices to be computed exactly. The resulting functional form is of an identical
structure to that for the real Ginibre ensemble, allowing in particular the corresponding
generalized partition function to be written as a Pfaffian. This is considered in Section
3 and used to give the probability $p_{k,N}$ that an $N \times N$ ($N$ even) member of the
interpolating ensemble has exactly $k$ real eigenvalues as a determinant of size $N/2$.
The Pfaffian formula for the generalized partition function implies the $k$-point
correlation functions can be written as a $2k \times 2k$ Pfaffian, with entries given in terms
of skew orthogonal polynomials. The latter are with respect to the skew inner product
corresponding to the entries of the Pfaffian. The main technical task is the computation
of these skew orthogonal polynomials. In Section 4 simple expressions in terms of Hermite
polynomials are obtained. The final section, Section 5, is concerned with
various scaled limits of the real-real and complex-complex correlations. 
For fixed $\tau$ the forms obtained are identical to the 
$\tau = 0$ case (real Ginibre ensemble) except for
a simple scaling of the coordinates which accounts for the change in the two-dimensional
density. The case $\tau = 0$ has been previously studied in \cite{FN07,So07} and most
comprehensively in the work of Borodin and Sinclair \cite{BS08} (this latter work
treats too the general real-complex correlations). The weakly non-symmetric limit of the
correlations is also computed.

\section{Definition of the ensemble and the eigenvalue PDF}
\setcounter{equation}{0}

Let $S$ be an element of the Gaussian orthogonal ensemble of $N \times N$ real symmetric
matrices, and thus have PDF of its independent elements proportional to
$e^{- {\rm Tr} \, S^2/2}$ (equivalently, the diagonal elements have distribution
N[0,1] while the strictly upper triangular elements have distribution
N$[0,1/\sqrt{2}]$). Let $A$ be an element of the anti-symmetric Gaussian
orthogonal ensemble of real anti-symmetric matrices which has PDF of its independent
elements proportional to $ e^{{\rm Tr} \,  A^2/2}$ (each strictly upper triangular
element is thus independently distributed according to N$[0,1/\sqrt{2}]$). With
$0 < \tau < 1$ and $c := (1 - \tau)/(1 + \tau)$ define random matrices $X$ according to
\begin{equation}\label{3}
X =  {1 \over \sqrt{b} } (S + \sqrt{c} A).
\end{equation}
When $\tau = 0$ and $b=1$, $X = S + A$. In this case each element of $X$ is independently distributed as
a standard Gaussian N[0,1] and so $X$ is a member of the real Ginibre ensemble. When
$\tau = 1$ and $b=1$, $X = S$ and so $X$ is a member of the GOE. Thus
$X$ interpolates between the real Ginibre ensemble and the GOE
as the parameter $\tau$ is varied from 1 down to 0.

The probability measure associated with the matrices $S$ and $A$ is
\begin{equation}\label{3.a}
(2 \pi)^{- N/2} \pi^{- N(N-1)/2} e^{-( {\rm Tr} \, S^2 - {\rm Tr} \, A^2)/2} (dA) (d S).
\end{equation}
From (\ref{3}) we compute that
$$
(d X) = 2^{N(N-1)/2} (\sqrt{c})^{N(N-1)/2} (\sqrt{b})^{-N^2} (d S) (d A),
$$
and we observe too that $S$ and $A$ can be written in terms of $X$ and $X^T$.
Thus we can change variables in (\ref{3.a}) to obtain for the PDF of the matrices $X$
\begin{equation}\label{4}
A_{\tau,b} \exp \Big ( - {b \over 2 (1 - \tau) } \Big (
{\rm Tr} \, X X^T - \tau {\rm Tr} \, X^2 \Big ) \Big ),
\end{equation}
where
\begin{equation}\label{4.a}
A_{\tau,b} = (\sqrt{c})^{- N(N-1)/2} (\sqrt{b})^{N^2} (2 \pi) ^{- N^2/2}.
\end{equation}
We seek the eigenvalue PDF corresponding to (\ref{4}).

A fundamental point is that because $X$ is real, there is a non-zero probability
that the eigenvalue will be real, and furthermore all complex eigenvalues must occur
in complex conjugate pairs. Thus the eigenvalue PDF decomposes into a sum of PDFs
$P_{k, (N-k)/2}(\{\lambda_j\}_{j=1,\dots,k}; \{x_j \pm i y_j \}_{j=1,\dots,(N-k)/2};
\tau,b)$ corresponding to having $k$ real eigenvalues $\{\lambda_j\}_{j=1,\dots,k}$ and
$(N-k)/2$ complex conjugate pairs of eigenvalues $ \{x_j \pm i y_j \}_{j=1,\dots,(N-k)/2}$
(for this to be non-zero $k$ and $N$ must have the same parity, a condition  which
will henceforth be assumed). In the case $\tau = 0$ and $b=1$ the probability
$P_{k, (N-k)/2}$ has been computed explicitly in \cite{LS91,Ed95} to give
\begin{eqnarray}\label{5}
&&
P_{k, (N-k)/2}(\{\lambda_j\}_{j=1,\dots,k}; \{x_j \pm i y_j \}_{j=1,\dots,(N-k)/2};
0,1)  \nonumber \\
&& =
{1 \over 2^{N(N+1)/4} \prod_{l=1}^N \Gamma(l/2) }
{2^{(N-k)/2} \over k! ((N-k)/2)! } \Big | \Delta(\{\lambda_l\}_{l=1,\dots,k} \cup
\{ x_j \pm i y_j \}_{j=1,\dots,(N-k)/2}) \Big | \nonumber \\
&& \qquad \times
e^{- \sum_{j=1}^k \lambda_j^2/2} e^{\sum_{j=1}^{(N-k)/2}(y_j^2 - x_j^2)}
\prod_{j=1}^{(N-k)/2} {\rm erfc}(\sqrt{2} y_j)
\end{eqnarray}
where $\Delta(\{z_p\}_{p=1,\dots,m}) := \prod_{j < l}^m (z_l - z_j)$.
In fact the eigenvalue PDF for general $\tau$ and $b$ is closely related to this functional form.

In the case $\tau = 0$ and $b=1$ we read off from (\ref{4}) that the PDF of the elements of
$X$ is given by $A_{0,1} e^{ - {\rm Tr} \, X X^T / 2}$. With $X \mapsto
\sqrt{b} X/ (1 - \tau)^{1/2}$ this latter PDF becomes
\begin{equation}\label{6A}
A_{0,1} b^{N^2/2} (1 - \tau)^{- N^2/2} e^{ - b {\rm Tr} \, X X^T / 2 (1 - \tau)},
\end{equation}
while the eigenvalue PDF is obtained from (\ref{5}) by a simple scaling
and so is equal to
\begin{eqnarray*}
&& b^{N/2} (1 - \tau)^{- N/2} P_{k, (N-k)/2}(\{ \sqrt{b} \lambda_j/(1 - \tau)^{1/2}\}_{j=1,\dots,k};
\nonumber \\
&& \quad  
\{\sqrt{b} x_j/(1 - \tau)^{1/2} \pm i \sqrt{b}
y_j/(1 - \tau)^{1/2} \}_{j=1,\dots,(N-k)/2};
0,1).
\end{eqnarray*}
Now (\ref{6A}) is a factor in (\ref{4}) while the remaining factor, proportional to
$ \exp \Big ( {\tau b \over 2 (1 - \tau) } {\rm Tr} \, X^2 \Big )$ can immediately be
written in terms of the eigenvalues of $X$. It follows from these considerations
that \cite{LS91}
\begin{eqnarray}\label{5.1}
&&
P_{k, (N-k)/2}(\{\lambda_j\}_{j=1,\dots,k}; \{x_j \pm i y_j \}_{j=1,\dots,(N-k)/2};
\tau,b)  \nonumber \\
&&
= {A_{\tau,b} \over A_{0,1}} (1 - \tau)^{N(N-1)/2}
\exp \Big (  {\tau b \over 2 (1 - \tau) } \Big (
\sum_{j=1}^k \lambda_j^2 + 2 \sum_{j=1}^{(N-k)/2} (x_j^2 - y_j^2) \Big ) \Big ) 
 P_{k, (N-k)/2} \nonumber \\
&&
\quad  (\{\sqrt{b} \lambda_j/(1 - \tau)^{1/2}\}_{j=1,\dots,k}; 
\{\sqrt{b} x_j/(1 - \tau)^{1/2} \pm i \sqrt{b} y_j/(1 - \tau)^{1/2} \}_{j=1,\dots,(N-k)/2};
0,1) \nonumber \\
&&
= 
{(\sqrt{b})^{N(N+1)/2}
(\sqrt{1 + \tau} )^{N(N-1)/2} \over 2^{N(N+1)/4} \prod_{l=1}^N \Gamma(l/2) }
{2^{(N-k)/2} \over k! ((N-k)/2)! } \Big | \Delta(\{\lambda_l\}_{l=1,\dots,k} \cup
\{ x_j \pm i y_j \}_{j=1,\dots,(N-k)/2}) \Big | \nonumber \\
&& \qquad \times
e^{- b \sum_{j=1}^k \lambda_j^2/2} e^{b \sum_{j=1}^{(N-k)/2}(y_j^2 - x_j^2)}
\prod_{j=1}^{(N-k)/2} {\rm erfc}\Big (\sqrt{2b \over 1 - \tau} y_j \Big ).
\end{eqnarray} 

Integrating $P_{k,(N-k)/2}$ over $\lambda_j \in \mathbb R$ $(j=1,\dots,k)$ and
$(x_j, y_j) \in \mathbb R_2^+$ $(j=1,\dots,(N-k)/2)$, where $\mathbb R_2^+ :=
\{ (x,y) \in \mathbb R^2: y > 0 \}$ gives the probability $p_{k,N}$ say that a
matrix of the form (\ref{3}) has exactly $k$ real eigenvalues.
The parameter $b$ then scales out of the problem and so for convenience may be
set equal to unity.
A discussion of a systematic approach to the calculation of these probabilities
is given in Section 3 below. The case $k=N$, when all eigenvalues are real, is
special and can be considered immediately. Thus, comparing (\ref{5}) and (\ref{5.1})
one sees that
$$
P_{N,0}(\{\lambda_j\}_{j=1,\dots,N};\tau,1) = (\sqrt{b})^{N^2/2}
(\sqrt{1 + \tau})^{N(N-1)/2} P_{N,0}(\{\lambda_j\}_{j=1,\dots,N};0,1).
$$
But we know from \cite{Ed95} that $p_{N,N} |_{\tau = 0 \atop b=1} = 2^{- N(N-1)/4}$ and so for
general $0 \le \tau \le 1$
\begin{equation}
p_{N,N} = \Big ( {2 \over 1 + \tau } \Big )^{- N(N-1)/4}.
\end{equation}

\section{Generalized partition function and the probabilities $p_{k,N}$}
\setcounter{equation}{0}

The generalized partition function associated with the PDF (\ref{5.1}) is defined by
\begin{eqnarray}\label{6}
&& Z_{k,(N-k)/2}[u,v] = \int_{-\infty}^\infty d\lambda_1 \cdots
 \int_{-\infty}^\infty d\lambda_k \, \prod_{l=1}^k u(\lambda_l)
\int_{\mathbb R_+^2} dx_1 dy_1 \cdots \int_{\mathbb R_+^2} dx_{(N-k)/2} dy_{(N-k)/2}
\nonumber \\
&&
\quad \times \prod_{l=1}^{(N-k)/2} v(x_l,y_l) P_{k,N-k}(\{\lambda_j\}_{j=1,\dots,k}; 
\{x_j \pm i y_j \}_{j=1,\dots,(N-k)/2};
\tau,b).
\end{eqnarray}
In view of the functional form (\ref{5.1}) this is structurally identical to the
case $\tau = 0$ and $b=1$, when a Pfaffian formula is known \cite{Si06}. Consequently
(\ref{6}) too has a Pfaffian evaluation.

\begin{prop}\label{pp1}
Let $\{p_{l-1}(x) \}_{l=1,\dots,N}$ be a set of monic polynomials of the indexed degree,
and let
\begin{eqnarray}\label{Za}
&&\alpha_{j,k}[u]  =  \int_{-\infty}^\infty dx \, u(x)  \int_{-\infty}^\infty dy \, u(y) 
e^{-b(x^2+y^2)/2}
p_{j-1}(x) p_{k-1}(y) {\rm sgn}(y-x) \label{6.4} \\
&&\beta_{j,k}[v]  =
 2 i \int_{{\mathbb R}_+^2} dx dy \, v(x,y) e^{b(y^2 - x^2)}
{\rm erfc} \Big (\sqrt{2 b \over 1 - \tau}   y \Big ) \nonumber \\ &&
\quad \times \Big ( p_{j-1}(x+iy) 
p_{k-1}({x-iy}) - p_{k-1}(x+iy) p_{j-1}(x-iy) \Big ). 
\end{eqnarray}
For $k, N$ even we have
\begin{equation}\label{Zb}
Z_{k,(N-k)/2}[u,v] = {(\sqrt{1 + \tau})^{N(N-1)/2} \over
2^{N(N+1)/4} \prod_{l=1}^N \Gamma(l/2) }
[ \zeta^{k/2}] {\rm Pf} \, [ \zeta \alpha_{j,l}[u] + \beta_{j,l}[v] ]_{j,l=1,\dots,N},
\end{equation}
where $[\zeta^p] f(\zeta)$ denotes the coefficient of $\zeta^p$ in $f(\zeta)$.
\end{prop}

We remark that the Pfaffian operation applies to even-dimensional anti-symmetric matrices;
the formula (\ref{Zb}) therefore requires modification for $N$ odd. Such a modification is
known \cite{Si06}, but to avoid having to consider separately the cases $N$ even and
$N$ odd, only the case $N$ even will be considered hereforth (it is planned to address
the case $N$ odd in a separate publication).

From the definitions, $Z_{k,(N-k)/2}[1,1] = p_{k,N}$ and so we have
\begin{equation}\label{A0}
p_{k,N} = {(\sqrt{1 + \tau} )^{N(N-1)/2} \over 2^{N(N+1)/4} \prod_{l=1}^N \Gamma(l/2)}
[\zeta^{k/2} ] {\rm Pf} \, [ \zeta \alpha_{j,l}[1] + \beta_{j,l}[1] ]_{j,l=1,\dots,N}
\Big |_{b=1}
\end{equation}
(it is valid to set $b=1$ since as noted below (\ref{5.1}) $p_{k,N}$ is independent
of $b$).
Suppose for definiteness that in (\ref{Za}) we choose $p_j(x) = x^j$
$(j=0,\dots,N-1)$. Changing variables $x \mapsto - x$, $y \mapsto - y$ shows
\begin{equation}\label{A1}
\alpha_{2j,2k}[1] = \alpha_{2j-1,2k-1}[1] = 0,
\end{equation}
while we see by introducing polar coordinates and changing variables
$\theta \mapsto \pi - \theta$ that furthermore
\begin{equation}\label{A2}
\beta_{2j,2k}[1] = \beta_{2j-1,2k-1}[1] = 0.
\end{equation}
The equations (\ref{A1}) and (\ref{A2}) give that the matrix in 
(\ref{A0}) has a checkerboard pattern of zeros. Recalling the general formula
$({\rm Pf} \, A)^2 = \det A$, rearranging rows and columns in $A$ so that all
non-zero entries are in the top right and bottom right $N \times N$ blocks,
and using the fact that the entries are anti-symmetric shows the Pfaffian can
be written as a determinant of half its size. Thus
\begin{equation}\label{tn}
p_{k,N} = {(\sqrt{1 + \tau} )^{N(N-1)/2} \over 2^{N(N+1)/4} \prod_{l=1}^N \Gamma(l/2)}
[\zeta^{k/2} ] \det [ \zeta \alpha_{2j-1,2k}[1] + \beta_{2j-1,2k}[1] ]_{j,k=1,\dots,N/2}
\end{equation}
where the quantities in the determinant have $b=1$.
It remains to evaluate these quantities.

In relation to $ \alpha_{2j-1,2k}[1]$, integration by parts shows
$$
\alpha_{2j-1,2k}[1] = 2(k-1) \alpha_{2j-1,2k-2}[1] + 2 \Gamma(j+k-3/2)
$$
and thus we obtain the explicit formula
\begin{equation}
\alpha_{2j-1,2k}[1] = 2^k (k-1)! \sum_{p=1}^k {\Gamma(j+p-3/2) \over 2^{p-1} (p-1)!}.
\end{equation}
For $\beta_{2j-1,2k}[1]$ we see from the definition that
\begin{eqnarray}\label{8}
&&\beta_{2j-1,2k}[1] = - 4 {\rm Im} \int_{\mathbb R^+} dx dy \,
e^{y^2 - x^2} {\rm erfc} \Big ( \sqrt{2 \over 1 - \tau} y \Big ) (x + i y)^{2j-2}
(x - i y)^{2k - 1} \nonumber \\
&& \qquad = - 4
\mathop{\sum_{l=0}^{2j - 2} \sum_{p=0}^{2k-1}}\limits_{l+p \: {\rm odd}}
\Big ( {2j - 2 \over l} \Big ) \Big ( {2k - 1 \over p} \Big ) (-1)^p
\Gamma(j+k-1 - (l+p)/2) I_{l+p}
\end{eqnarray}
where
$$
I_j := \int_0^\infty y^j {\rm erfc} \Big ( \sqrt{2 \over 1 - \tau} \Big ) e^{y^2} \, dy
\qquad (j \: \: {\rm odd}).
$$
Integration by parts shows
$$
I_j = - {(j-1) \over 2} I_{j-2} + {1 \over \sqrt{\pi} }
\sqrt{2 \over 1 - \tau} \Big ( {1 - \tau \over 1 + \tau} \Big )^{j/2}
{1 \over 2} \Gamma(j/2) - {1 \over 2} \delta_{j,1}.
$$
This recurrence has solution
\begin{equation}\label{3.11}
I_j = {(-1)^{(j-1)/2} ((j-1)/2)! \over 2}
\Big ( \sqrt{2 \over 1 + \tau}
\sum_{p=0}^{(j-1)/2} (-1)^p \Big ( {1 - \tau \over 1 + \tau} \Big )^p
{(1/2)_p \over p!} - 1 \Big )
\end{equation}
which makes $\beta_{2j-1,2k}[1]$ explicit. In the case $\tau = 0$ (\ref{3.11})
reduces to a result of
Edelman \cite{Ed95}. 

In the case $\tau = 0$ explicit formulas for $p_{k,N}$ were first calculated in
\cite{Ed95}, by direct integration of (\ref{5}). A formula closely related to
(\ref{A0}) was given in \cite{KA05}, while the formula (\ref{A0}) itself in the
case $\tau = 0$ was given in \cite{FN07}. A determinant formula equivalent to
(\ref{tn}) in the case $\tau = 0$, but derived using a different strategy,
is given in \cite{AK07}.

\section{Correlations and skew orthogonal polynomials}
\setcounter{equation}{0}
As realizations of $X$ are not conditioned on the number of real eigenvalues $k$,
the generalized partition function $Z_N[u,v]$ appropriate for calculation of the correlation
functions is obtained by summing (\ref{6}) over all allowed $k$,
$$
Z_N[u,v] = \sum_{k=0 \atop k \: {\rm even}}^N Z_{k,(N-k)/2} [ u,v].
$$
It follows from (\ref{Zb}) that
\begin{equation}\label{Zc}
Z_{N}[u,v] = {(\sqrt{1 + \tau})^{N(N-1)/2} \over
2^{N(N+1)/4} \prod_{l=1}^N \Gamma(l/2) }
 {\rm Pf} \, [ \zeta \alpha_{j,l}[u] + \beta_{j,l}[v] ]_{j,l=1,\dots,N}.
\end{equation}
Correlation functions can be calculated from this by functional differentiation.
For example, the $n$-point correlation function between real eigenvalues at
$x_1,\dots,x_n$ is given by
\begin{equation}\label{4.22a}
\rho_{(n)}^{\rm r}(x_1,\dots,x_n) = {1 \over Z_N[1,1] }
{\delta^n \over \delta u(x_1) \cdots \delta u(x_n) }
Z_N[u,1] \Big |_{u=1}.
\end{equation}

Moreover, all $n$-point correlation functions can be expressed in terms of
a $2k \times 2k$ Pfaffian \cite{BS08} (see also \cite{FN07} in the case of all
real eigenvalues in the correlation, or all complex eigenvalues). The structure
of these formulas is the same for all allowed $\tau$ and $b$, which is a consequence of the
structure of the entries of (\ref{Zc}) being the same for all allowed $\tau$
and $b$.
In particular, in the case (\ref{4.22a})
\begin{equation}\label{rr}
\rho_{(n)}^{\rm r}(x_1,\dots,x_n) =
{\rm Pf}
\left [ \begin{array}{cc} - \tilde{I}^{\rm r}(x_j,x_k) & S^{\rm r}(x_j,x_k) \\
- S^{\rm r}(x_k,x_j) & D^{\rm r}(x_j,x_k) \end{array} \right ],
\end{equation}
where with 
\begin{equation}\label{rr1}
\Phi_k(x) := \int_{-\infty}^\infty {\rm sgn}(x-y) p_k(y) e^{- y^2/2(1 + \tau)} \, dy
\end{equation}
one has
\begin{eqnarray}
&& S^{\rm r}(x,y)  = \sum_{k=0}^{N/2 - 1} {e^{-y^2/2(1 + \tau)} \over u_k} \Big (
\Phi_{2k}(x) p_{2k+1}(y) - \Phi_{2k+1}(x) p_{2k}(y) \Big ) \label{12.Sr} \\
&& D^{\rm r}(x,y) = {\partial \over \partial x} S^{\rm r}(x,y), \qquad
\tilde{I}^{\rm r}(x, y) = {1 \over 2} {\rm sgn}(y-x) - \int_x^y S^{\rm r}(x,z) \, dz. \label{12.Sr1}
\end{eqnarray}
We note too the explicit Pfaffian formula for the correlation between complex
eigenvalues
\begin{eqnarray}\label{St0}
&& \rho_{(n)}^{\rm c}((x_1,y_1),\dots,(x_n,y_n)) =
\prod_{l=1}^n \Big ( 2i e^{(y_l^2 - x_l^2)/(1 + \tau)}
{\rm erfc} \Big ( \sqrt{2 \over 1 - \tau^2} y_l \Big ) \Big ) \nonumber \\
&&
\qquad \times {\rm Pf}
\left [ \begin{array}{cc} S^{\rm c}_\tau(\bar{z}_j, \bar{z}_k) & S^{\rm c}_\tau(\bar{z}_j,z_k) \\
S^{\rm c}_\tau({z}_j, \bar{z}_k) & S^{\rm c}_\tau({z}_j,z_k) \end{array} \right ],
\end{eqnarray}
where $z_j := x_j + i y_j$ and
with $q_{2j-2}(z) := - p_{2j-1}(z)$, $q_{2j-1}(z) := p_{2j-2}(z)$,
\begin{equation}\label{St}
S^{\rm c}_\tau(w,z) = \sum_{j=1}^N {p_{j-1}(w) q_{j-1}(z) \over u_{[(j-1)/2]} }.
\end{equation}
The dependence on $\tau$ comes in through the requirements of the polynomials
$\{p_j(x)\}_{j=0,1,\dots}$. In addition to being monic of the appropriate degree
as in Proposition \ref{pp1}, they must be skew orthogonal with respect to the
skew inner product associated with the matrix in (\ref{Zc}) for $u=v=1$.

Explicitly, this inner product reads
\begin{equation}\label{fg}
(f,g) := (f,g)_{\rm r} + (f,g)_{\rm c}
\end{equation}
with
\begin{eqnarray*}
&& (f,g)_{\rm r} :=
\int_{-\infty}^\infty dx   \int_{-\infty}^\infty dy  e^{-b (x^2+y^2)/2}
f(x) g(y) {\rm sgn}(y-x)  \\
&&(f,g)_{\rm c}  :=
 2 i \int_{{\mathbb R}_+^2} dx dy \,  e^{b(y^2 - x^2)}
{\rm erfc} \Big (\sqrt{2 b \over 1 - \tau}   y \Big ) \Big ( f(x+iy)
g(x-iy) - g(x+iy) f(x-iy) \Big ).
\end{eqnarray*}
The set $\{p_j(x)\}_{j=0,1,\dots}$ is said to be skew orthogonal if
$$
(p_{2j}, p_{2k}) = (p_{2j+1}, p_{2k+1}) = 0 \: \: (j,k=0,1,\dots) \qquad
(p_{2j},p_{2k+1}) = 0 \: \: (j,k=0,1,\dots \: j \ne k),
$$
while $(p_{2j},p_{2j+1}) = u_{j} \ne 0$.
The main technical task then is to compute these polynomials. Note that the
parameter $b$ acts as a scale of the coordinates, and so there is no loss
of generality in setting $b$ to a specific value. It turns out that a convenient
choice is $b = 1/(1 + \tau)$. Making this choice, the skew inner product of
interest reads
\begin{equation}\label{fgb}
\langle f,g \rangle := \langle f,g  \rangle_{\rm r} + \langle f,g \rangle_{\rm c}
\end{equation}
with
\begin{eqnarray*}
&&\langle f,g \rangle_{\rm r} := \int_{-\infty}^{\infty} {\rm d}x
\int_{-\infty}^{\infty} {\rm d}y {\rm e}^{-\frac{x^2 + y^2}{2(1 + \tau)}}
f(x) g(y) {\rm sgn}(y-x), \nonumber \\
&& \langle f,g \rangle_{\rm c}  :=  2 i \int_{-\infty}^{\infty} {\rm d}x
\int_0^{\infty} {\rm d}y  {\rm e}^{\frac{y^2-x^2}{1 + \tau}}
{\rm erfc}\left(\sqrt{\frac{2}{1 - \tau^2}} \ y \right)
\left[ f(x + i y) g(x - i y) - g(x + i y) f(x - i y) \right].
\end{eqnarray*}

\begin{thm}\label{hm1}
Introduce the scaled monic Hermite polynomials
\begin{equation}\label{CH}
C_n(z) = \Big ( {\tau \over 2} \Big )^{n/2} H_n \Big ( {z \over \sqrt{2 \tau} } \Big ).
\end{equation}
The family of monic polynomials $\{ R_j(z) \}_{j=0,1,\dots}$ with
\begin{equation}\label{RC}
R_{2n+1}(z) = C_{2n+1}(z) - 2n C_{2n-1}(z), \qquad R_{2n}(z) = C_{2n}(z)
\end{equation}
are skew orthogonal with respect to the skew inner product (\ref{fgb}).
Furthermore, the normalization $r_n$ is given by
\begin{equation}\label{RC1}
r_n := \langle R_{2n}, R_{2n+1} \rangle = (2n)! 2 \sqrt{2 \pi} (1 + \tau).
\end{equation}
\end{thm}

\noindent
{\it Proof.} \quad Note that $R_{2n+1}(z)$ is an odd polynomial while $R_{2n}(z)$ is
even. These properties are sufficient for the derivation of (\ref{A1}) and (\ref{A2})
so it is immediate that
\begin{equation}\label{F1}
\langle R_{2j}, R_{2k} \rangle = \langle R_{2j+1}, R_{2k+1}  \rangle = 0.
\end{equation}
It remains to verify that
\begin{equation}\label{F2}
\langle R_{2j+1}, R_{2k} \rangle = 0
\end{equation}
for $k \ne j$, and that for $k=j$ the normalization (\ref{RC1}) results. This will
be done by computing the explicit form of the skew inner product between $C_{2j+1}$
and $C_{2k}$,
\begin{equation}
\label{m1}
\langle C_{2 j + 1}, C_{2 k} \rangle 
= \left\{ \begin{array}{ll} \displaystyle 
- 2^{j + k + (3/2)} j! \Gamma\left( k + \frac{1}{2} \right) (1 + \tau), 
&  j \geq k, \\ 
0, & j < k. \end{array} \right. 
\end{equation}
Assuming (\ref{m1}), the explicit formulas (\ref{RC}) show that (\ref{F2}) is valid
and furthermore give the normalization (\ref{RC1}). The task is thus reduced to
proving (\ref{m1}).
For this, repeated use will be made of the properties of the Hermite polynomials
(\ref{CH}) 
\begin{equation}\label{CH.4}
\frac{{\rm d}}{{\rm d} z} C_n(z) = n C_{n-1}(z),
\end{equation}
\begin{equation}
\label{ztimes}
z C_n(z) = C_{n+1}(z) + n \tau C_{n-1}(z).
\end{equation}

Let us first define    
\begin{equation}
I_{j,k} = \int_{-\infty}^{\infty} {\rm d}x 
\int_{-\infty}^{\infty} {\rm d}y \, {\rm e}^{-\frac{x^2 + y^2}{2(1 + \tau)}} 
C_{2 j + 1}(x) 
C_{2 k}(y) {\rm sgn}(y-x). 
\end{equation}
Then a partial integration over $x$ gives
\begin{eqnarray}
& & I_{j,k} \nonumber \\ 
&  &  \quad = \int_{-\infty}^{\infty} 
{\rm d}y {\rm e}^{-\frac{y^2}{2(1 + \tau)}} C_{2 k}(y) 
\left[ {\rm e}^{-\frac{x^2}{2(1 + \tau)}} 
\frac{C_{2 j + 2}(x)}{2 j + 2} {\rm sgn}(y-x) 
\right]_{x = -\infty}^{x = \infty} \nonumber \\ 
&  & \qquad + \frac{1}{1 + \tau} \int_{-\infty}^{\infty} {\rm d}x 
\int_{-\infty}^{\infty} {\rm d}y {\rm e}^{-\frac{x^2 + y^2}{2(1 + \tau)}} 
\frac{x C_{2 j + 2}(x)}{2 j + 2} 
C_{2 k}(y) {\rm sgn}(y-x) \nonumber \\  
&  & \qquad + \int_{-\infty}^{\infty} {\rm d}x 
\int_{-\infty}^{\infty} {\rm d}y {\rm e}^{-\frac{x^2 + y^2}{2(1 + \tau)}} 
\frac{C_{2 j + 2}(x)}{2 j + 2} 
C_{2 k}(y) 2 \delta(y-x) \nonumber \\ 
&  & \quad = \frac{1}{1 + \tau} \frac{1}{2 j + 2} \int_{-\infty}^{\infty} {\rm d}x 
\int_{-\infty}^{\infty} {\rm d}y {\rm e}^{-\frac{x^2 + y^2}{2(1 + \tau)}} 
C_{2 j + 3}(x) C_{2 k}(y) {\rm sgn}(y-x) \nonumber \\  
&  & \qquad + \frac{\tau}{1 + \tau} \int_{-\infty}^{\infty} {\rm d}x 
\int_{-\infty}^{\infty} {\rm d}y {\rm e}^{-\frac{x^2 + y^2}{2(1 + \tau)}} 
C_{2 j + 1}(x) C_{2 k}(y) {\rm sgn}(y-x) \nonumber \\  
&  & \qquad + \frac{2}{2 j + 2} \int_{-\infty}^{\infty} {\rm d}x 
{\rm e}^{-\frac{x^2}{1 + \tau}} 
C_{2 j + 2}(x) C_{2 k}(x) \nonumber \\ 
&  & \quad = \frac{1}{1 + \tau} \frac{1}{2 j + 2} I_{j+1,k} + \frac{\tau}{1 + \tau} 
I_{j,k} + \frac{2}{2 j + 2} \frac{\xi_{j,k}}{1 + \tau},
\end{eqnarray}
where
\begin{equation}
\xi_{j,k} = (1 + \tau) 
\int_{-\infty}^{\infty} {\rm d}x {\rm e}^{-\frac{x^2}{1 + \tau}} 
C_{2 j + 2}(x) C_{2 k}(x).
\end{equation}
Similarly a partial integration over $y$ gives
\begin{eqnarray}
& & I_{j,k} \nonumber \\ 
&  & \quad = \int_{-\infty}^{\infty} 
{\rm d}x {\rm e}^{-\frac{x^2}{2(1 + \tau)}} C_{2 j + 1}(x) \left[ 
{\rm e}^{-\frac{y^2}{2(1 + \tau)}} 
\frac{C_{2 k + 1}(y)}{2 k + 1} {\rm sgn}(y-x) 
\right]_{y = -\infty}^{y = \infty} \nonumber \\ 
&  & \qquad + \frac{1}{1 + \tau} 
\int_{-\infty}^{\infty} {\rm d}x \int_{-\infty}^{\infty} 
{\rm d}y {\rm e}^{-\frac{x^2 + y^2}{2(1 + \tau)}} 
C_{2 j + 1}(x) \frac{y C_{2 k + 1}(y)}{2 k + 1} 
{\rm sgn}(y-x) \nonumber \\  
&  & \qquad - \int_{-\infty}^{\infty} {\rm d}x 
\int_{-\infty}^{\infty} {\rm d}y {\rm e}^{-\frac{x^2 + y^2}{2(1 + \tau)}} 
C_{2 j + 1}(x) \frac{C_{2 k + 1}(y)}{2 k + 1} 2 
\delta(y-x) \nonumber \\ 
&  & \quad = \frac{1}{1 + \tau} \frac{1}{2 k + 1} \int_{-\infty}^{\infty} {\rm d}x 
\int_{-\infty}^{\infty} {\rm d}y {\rm e}^{-\frac{x^2 + y^2}{2(1 + \tau)}} 
C_{2 j + 1}(x) C_{2 k + 2}(y) 
{\rm sgn}(y-x) \nonumber \\  
&  & \qquad \frac{\tau}{1 + \tau} \int_{-\infty}^{\infty} {\rm d}x 
\int_{-\infty}^{\infty} {\rm d}y {\rm e}^{-\frac{x^2 + y^2}{2(1 + \tau)}} 
C_{2 j + 1}(x) C_{2 k}(y) 
{\rm sgn}(y-x) \nonumber \\  
&  & \qquad - \frac{2}{2 k + 1} \int_{-\infty}^{\infty} {\rm d}x 
{\rm e}^{-\frac{x^2}{1 + \tau}} 
C_{2 j + 1}(x) C_{2 k + 1}(x) \nonumber \\ 
&  & \quad = \frac{1}{1 + \tau} 
\frac{1}{2 k + 1} I_{j,k + 1} + \frac{\tau}{1 + \tau} I_{j,k} 
- \frac{2}{2 k + 1}\frac{\eta_{j,k}}{1 + \tau}, 
\end{eqnarray}
where
\begin{equation}
\eta_{j,k} = (1 + \tau) \int_{-\infty}^{\infty} {\rm d}x 
{\rm e}^{-\frac{x^2}{1 + \tau}} 
C_{2 j + 1}(x) C_{2 k + 1}(x).
\end{equation}
Thus we obtain recursion relations
\begin{eqnarray}
\label{rec1}
I_{j+1,k} & = & (2 j + 2) I_{j,k} - 2 \xi_{j,k}, \nonumber \\   
I_{j,k+1} & = & (2 k + 1) I_{j,k} + 2 \eta_{j,k}.
\end{eqnarray}
\par
Let us next derive the recursion relations for 
\begin{eqnarray}
&&J_{j,k}  =  2 i \int_{-\infty}^{\infty} {\rm d}x 
\int_0^{\infty} {\rm d}y  {\rm e}^{\frac{y^2 - x^2}{1 + \tau}} 
{\rm erfc}(\gamma y) 
\nonumber \\ 
&  & \qquad \times \left[ C_{2 j + 1}(x + i y) C_{2 k}(x - i y) 
- C_{2 k}(x + i y) C_{2 j + 1}(x - i y) \right]
\end{eqnarray}
with 
\begin{equation}
\gamma = \sqrt{\frac{2}{1 - \tau^2}}.
\end{equation}
For that purpose, we consider an integral   
\begin{eqnarray}
&& K_{j,k}  =  2 i \int_{-\infty}^{\infty} {\rm d}x 
\int_0^{\infty} {\rm d}y  {\rm e}^{\frac{y^2 - x^2}{1 + \tau}} 
{\rm erfc}(\gamma y) 
\nonumber \\ 
& & \times \qquad  \left[ C_{2 j + 2}(x + i y) C_{2 k - 1}(x - i y)
- C_{2 k - 1}(x + i y) C_{2 j + 2}(x - i y) \right].
\end{eqnarray}
For $k \geq 1$, a partial integration over $x$ gives
\begin{eqnarray}
\label{kjk1}
& & K_{j,k} \nonumber \\ 
&  & \quad = 2 i \int_0^{\infty}{\rm d}y {\rm e}^{\frac{y^2}{1 + \tau}} 
{\rm erfc}(\gamma y)  
\left[ {\rm e}^{-\frac{x^2}{1 + \tau}} C_{2 j + 2}(x + i y) 
\frac{C_{2 k}(x - i y)}{2 k} \right]_{x = - 
\infty}^{x = \infty} \nonumber \\ 
&  & \qquad + 4 i \frac{1}{1 + \tau} \int_{-\infty}^{\infty} {\rm d}x 
\int_0^{\infty} {\rm d}y  {\rm e}^{\frac{y^2-x^2}{1 + \tau}} 
{\rm erfc}(\gamma y) 
x C_{2 j + 2}(x + i y) \frac{C_{2 k}(x - i y)}{2 k} \nonumber \\ 
&   & \qquad - 2 i \int_{-\infty}^{\infty} {\rm d}x 
\int_0^{\infty} {\rm d}y  {\rm e}^{\frac{y^2-x^2}{1 + \tau}} 
{\rm erfc}(\gamma y) 
(2 j + 2) C_{2 j + 1}(x + i y) \frac{C_{2 k}(x - i y)}{2 k} \nonumber \\ 
&  & \qquad + c.c. \nonumber \\  
&  & \quad = \frac{2 i}{k} \frac{1}{1 + \tau} \int_{-\infty}^{\infty} {\rm d}x 
\int_0^{\infty} {\rm d}y  {\rm e}^{\frac{y^2-x^2}{1 + \tau}} 
{\rm erfc}(\gamma y) 
x C_{2 j + 2}(x + i y) C_{2 k}(x - i y) \nonumber \\ 
&  & \qquad - 2 i \frac{j + 1}{k} \int_{-\infty}^{\infty} {\rm d}x 
\int_0^{\infty} {\rm d}y  {\rm e}^{\frac{y^2-x^2}{1 + \tau}} 
{\rm erfc}(\gamma y) C_{2 j + 1}(x + i y) C_{2 k}(x - i y) \nonumber \\ 
&  & \qquad + c.c. 
\end{eqnarray}
On the other hand, a partial integration over $y$ gives
\begin{eqnarray}
\label{kjk2}
& & K_{j,k} \nonumber \\ 
&  & \quad = 2 i \int_{-\infty}^{\infty} {\rm d}x 
{\rm e}^{-\frac{x^2}{1 + \tau}} \left[ {\rm e}^{\frac{y^2}{1 + \tau}} 
{\rm erfc}(\gamma y) 
C_{2 j + 2}(x + i y) \frac{i C_{2 k}(x - i y)}{2 k} 
\right]_{y = 0}^{y = \infty} \nonumber \\   
&  & \qquad - 4 i \frac{1}{1 + \tau} \int_{-\infty}^{\infty} {\rm d}x 
\int_0^{\infty} {\rm d}y  {\rm e}^{\frac{y^2-x^2}{1 + \tau}} 
{\rm erfc}(\gamma y) 
y C_{2 j + 2}(x + i y) \frac{i C_{2 k}(x - i y)}{2 k} \nonumber \\ 
&  & \qquad + 2 i  \int_{-\infty}^{\infty} {\rm d}x 
\int_0^{\infty} {\rm d}y  {\rm e}^{\frac{y^2-x^2}{1 + \tau}} 
{\rm erfc}(\gamma y) 
(2 j + 2) C_{2 j + 1}(x + i y) 
\frac{C_{2 k}(x - i y)}{2 k} \nonumber \\  
&  & \qquad + 2 i \frac{2 \gamma}{\sqrt{\pi}} \int_{-\infty}^{\infty} {\rm d}x 
\int_0^{\infty} {\rm d}y  {\rm e}^{\frac{y^2-x^2}{1 + \tau}} 
{\rm e}^{-(\gamma y)^2}
C_{2 j + 2}(x + i y) \frac{i C_{2 k}(x - i y)}{2 k} + c.c. \nonumber \\  
&  & \quad + \frac{1}{k} \int_{-\infty}^{\infty} {\rm d}x 
{\rm e}^{-\frac{x^2}{1 + \tau}} C_{2 j + 2}(x) C_{2 k}(x) \nonumber \\   
&   & \qquad - \frac{2 i}{k} \frac{1}{1 + \tau} 
\int_{-\infty}^{\infty} {\rm d}x 
\int_0^{\infty} {\rm d}y  {\rm e}^{\frac{y^2-x^2}{1 + \tau}} 
{\rm erfc}(\gamma y) 
i y C_{2 j + 2}(x + i y) C_{2 k}(x - i y) \nonumber \\ 
&  & \quad + 2 i \frac{j+1}{k} \int_{-\infty}^{\infty} {\rm d}x 
\int_0^{\infty} {\rm d}y  {\rm e}^{\frac{y^2-x^2}{1 + \tau}} 
{\rm erfc}(\gamma y) C_{2 j + 1}(x + i y) C_{2 k}(x - i y) \nonumber \\  
&  & \quad - \frac{2}{k}  \sqrt{\frac{2}{\pi(1 - \tau^2)}} 
\int_{-\infty}^{\infty} {\rm d}x 
\int_0^{\infty} {\rm d}y  {\rm e}^{-\frac{x^2}{1 + \tau} - 
\frac{y^2}{1 - \tau}} 
C_{2 j + 2}(x + i y) C_{2 k}(x - i y) \nonumber \\ 
&  & \qquad + c.c. 
\end{eqnarray}
Comparing (\ref{kjk1}) and ({\ref{kjk2}), we obtain
\begin{eqnarray}
& & \frac{2 i}{k} \frac{1}{1 + \tau} \int_{-\infty}^{\infty} {\rm d}x 
\int_0^{\infty} {\rm d}y  {\rm e}^{\frac{y^2 - x^2}{1 + \tau}} 
{\rm erfc}(\gamma y) 
(x + i y) C_{2 j + 2}(x + i y) C_{2 k}(x - i y) \nonumber \\ 
&  & - 4 i \frac{j + 1}{k} \int_{-\infty}^{\infty} {\rm d}x 
\int_0^{\infty} {\rm d}y  {\rm e}^{\frac{y^2-x^2}{1 + \tau}} 
{\rm erfc}(\gamma y) C_{2 j + 1}(x + i y) C_{2 k}(x - i y)^{2 k} 
\nonumber \\ &  & \qquad + c.c. 
\nonumber \\ 
&  & = \frac{2}{k} \int_{-\infty}^{\infty} {\rm d}x 
{\rm e}^{-\frac{x^2}{1 + \tau}} C_{2 j + 2}(x) C_{2 k}(x) 
\nonumber \\ 
&  & \quad - \frac{2}{k} \sqrt{\frac{2}{\pi(1 - \tau^2)}} 
\int_{-\infty}^{\infty} {\rm d}x 
\int_0^{\infty} {\rm d}y  {\rm e}^{-\frac{x^2}{1 + \tau} - 
\frac{y^2}{1 - \tau}} 
\nonumber \\ 
&  &  \quad \times
\left[ C_{2 j + 2}(x + i y) C_{2 k}(x - i y) +   
C_{2 j + 2}(x - i y) C_{2 k}(x + i y) \right].
\end{eqnarray}
Therefore, noting (\ref{ztimes}) and the 
orthogonality relation (see e.g.~\cite{DGIL94})
\begin{equation}
\int_{-\infty}^{\infty} {\rm d}x 
\int_{-\infty}^{\infty} {\rm d}y 
{\rm e}^{-\frac{x^2}{1 + \tau}} {\rm e}^{-\frac{y^2}{1 - \tau}} 
C_m(x + i y) C_n(x - i y) = \pi m! \sqrt{1 - \tau^2} \delta_{m,n},
\end{equation}
we can derive
\begin{equation}
\label{rec2}
J_{j+1,k} = (2 j + 2) J_{j,k} + 2 \xi_{j,k} 
- 2 \sqrt{2 \pi} (2 j + 2)! \ (1 + \tau) \ \delta_{j+1,k}.   
\end{equation}
In order to derive another recursion relation, we 
similarly employ partial integrations to find 
\begin{eqnarray}
\label{kjk3}
& & K_{j-1,k+1} 
\nonumber \\ 
&  & \quad = 2 i \int_0^{\infty}{\rm d}y {\rm e}^{\frac{y^2}{1 + \tau}} 
{\rm erfc}(\gamma y)  
\left[ {\rm e}^{-\frac{x^2}{1 + \tau}} \frac{C_{2 j + 1}(x + i y)}{2 j + 1} 
C_{2 k + 1}(x - i y) \right]_{x = - 
\infty}^{x = \infty} \nonumber \\ 
&  & \qquad + 4 i \frac{1}{1 + \tau} \int_{-\infty}^{\infty} {\rm d}x 
\int_0^{\infty} {\rm d}y  {\rm e}^{\frac{y^2-x^2}{1 + \tau}} 
{\rm erfc}(\gamma y) 
\frac{C_{2 j + 1}(x + i y)}{2 j + 1} x C_{2 k + 1}(x - i y) \nonumber \\ 
&  & \qquad - 2 i \int_{-\infty}^{\infty} {\rm d}x 
\int_0^{\infty} {\rm d}y  {\rm e}^{\frac{y^2-x^2}{1 + \tau}} 
{\rm erfc}(\gamma y) 
\frac{C_{2 j + 1}(x + i y)}{2 j + 1} 
(2 k + 1) C_{2 k}(x - i y) 
\nonumber \\ 
&  & \qquad + c.c. \nonumber \\  
&  & \quad = \frac{4 i}{2 j + 1} \frac{1}{1 + \tau} 
\int_{-\infty}^{\infty} {\rm d}x 
\int_0^{\infty} {\rm d}y  {\rm e}^{\frac{y^2-x^2}{1 + \tau}} 
{\rm erfc}(\gamma y) 
C_{2 j + 1}(x + i y) x C_{2 k + 1}(x - i y) \nonumber \\ 
&  & \qquad - 2 i \frac{2 k + 1}{2 j + 1} \int_{-\infty}^{\infty} {\rm d}x 
\int_0^{\infty} {\rm d}y  {\rm e}^{\frac{y^2-x^2}{1 + \tau}} 
{\rm erfc}(\gamma y) C_{2 j + 1}(x + i y) 
C_{2 k}(x - i y) + c.c. \nonumber \\  
\nonumber \\ 
\end{eqnarray}
and
\begin{eqnarray}
\label{kjk4}
& & K_{j-1,k+1} 
\nonumber \\ 
&  & \quad = 2 i \int_{-\infty}^{\infty} {\rm d}x 
{\rm e}^{-\frac{x^2}{1 + \tau}} \left[ {\rm e}^{\frac{y^2}{1 + \tau}} 
{\rm erfc}(\gamma y) 
\frac{C_{2 j + 1}(x + i y)}{i (2 j + 1)} C_{2 k + 1}(x - i y) 
\right]_{y = 0}^{y = \infty} \nonumber \\   
&  & \qquad - 4 i \frac{1}{1 + \tau} \int_{-\infty}^{\infty} {\rm d}x 
\int_0^{\infty} {\rm d}y  {\rm e}^{\frac{y^2-x^2}{1 + \tau}} 
{\rm erfc}(\gamma y) 
\frac{C_{2 j + 1}(x + i y)}{i (2 j + 1)} y C_{2 k + 1}(x - i y) 
\nonumber \\ 
&  & \qquad + 2 i \int_{-\infty}^{\infty} {\rm d}x 
\int_0^{\infty} {\rm d}y  {\rm e}^{\frac{y^2-x^2}{1 + \tau}} 
{\rm erfc}(\gamma y) 
\frac{C_{2 j + 1}(x + i y)}{2 j + 1}  
(2 k + 1) C_{2 k}(x - i y) \nonumber \\  
&  & \qquad + 2 i \frac{2 \gamma}{\sqrt{\pi}} \int_{-\infty}^{\infty} {\rm d}x 
\int_0^{\infty} {\rm d}y  {\rm e}^{\frac{y^2-x^2}{1 + \tau}} 
{\rm e}^{-(\gamma y)^2} 
\frac{C_{2 j + 1}(x + i y)}{i (2 j + 1)} 
C_{2 k + 1}(x - i y) + c.c. \nonumber \\  
&  & \quad = - \frac{2}{2 j + 1} \int_{-\infty}^{\infty} {\rm d}x 
{\rm e}^{-\frac{x^2}{1 + \tau}} C_{2 j + 1}(x) C_{2 k + 1}(x) \nonumber \\   
&   & \qquad + \frac{4 i}{2 j + 1} \frac{1}{1 + \tau} 
\int_{-\infty}^{\infty} {\rm d}x 
\int_0^{\infty} {\rm d}y  {\rm e}^{\frac{y^2-x^2}{1 + \tau}} 
{\rm erfc}(\gamma y) 
C_{2 j + 1}(x + i y) i y C_{2 k + 1}(x - i y) \nonumber \\ 
&  & \qquad + 2 i \frac{2 k + 1}{2 j + 1} \int_{-\infty}^{\infty} {\rm d}x 
\int_0^{\infty} {\rm d}y  {\rm e}^{\frac{y^2-x^2}{1 + \tau}} 
{\rm erfc}(\gamma y) C_{2 j + 1}(x + i y) 
C_{2 k}(x - i y) \nonumber \\  
&  & \qquad + \frac{4}{2 j + 1} \sqrt{\frac{2}{\pi (1 - \tau^2)}} 
\int_{-\infty}^{\infty} {\rm d}x 
\int_0^{\infty} {\rm d}y  {\rm e}^{-\frac{x^2}{1 + \tau} 
- \frac{y^2}{1 - \tau}} 
C_{2 j + 1}(x + i y) C_{2 k + 1}(x - i y) 
\nonumber \\ 
&  & \quad + c.c. 
\end{eqnarray}
A comparison of (\ref{kjk3}) and ({\ref{kjk4}) yields
\begin{eqnarray}
& & \frac{4 i}{2 j + 1} \frac{1}{1 + \tau} 
\int_{-\infty}^{\infty} {\rm d}x 
\int_0^{\infty} {\rm d}y  
{\rm e}^{\frac{y^2-x^2}{1 + \tau}} 
{\rm erfc}(\gamma y) \nonumber \\ 
& & \quad \times 
C_{2 j + 1}(x + i y) (x - i y) C_{2 k + 1}(x - i y) \nonumber \\ 
&  & \quad - 4 i \frac{2 k + 1}{2 j + 1} \int_{-\infty}^{\infty} {\rm d}x 
\int_0^{\infty} {\rm d}y  {\rm e}^{\frac{y^2-x^2}{1 + \tau}} 
{\rm erfc}(\gamma y) C_{2 j + 1}(x + i y) C_{2 k}(x - i y) + c.c.  
\nonumber \\ 
&  & = - \frac{4}{2 j + 1} \int_{-\infty}^{\infty} {\rm d}x 
{\rm e}^{-\frac{x^2}{1 + \tau}} C_{2 j + 1}(x) C_{2 k + 1}(x) 
\nonumber \\ 
&  &  \quad +
\frac{4}{2 j + 1} \sqrt{\frac{2}{\pi (1 - \tau^2)}} 
\int_{-\infty}^{\infty} {\rm d}x 
\int_0^{\infty} {\rm d}y  {\rm e}^{-\frac{x^2}{1 + \tau} 
- \frac{y^2}{1 - \tau}} \nonumber \\ 
& & \quad \times   
\left[ C_{2 j + 1}(x + i y) C_{2 k + 1}(x - i y) +   
C_{2 j + 1}(x - i y) C_{2 k + 1}(x + i y) \right].
\end{eqnarray}    
As before it follows that
\begin{equation}
\label{rec3}
J_{j,k+1} = (2 k + 1) J_{j,k} - 2 \eta_{j,k}  
+ 2 \sqrt{2 \pi} (2 j + 1)! \ (1 + \tau) \ \delta_{j,k}.   
\end{equation}
\par
Let us employ the notation 
\begin{equation}
L_{j,k} = I_{j,k} + J_{j,k}.  
\end{equation}
Then, from (\ref{rec1}), (\ref{rec2}) and (\ref{rec3}), 
we obtain the recursion relations 
\begin{eqnarray}
L_{j+1,k} = (2 j + 2) L_{j,k} - 2 \sqrt{2 \pi} (2 j + 2)! \ 
(1 + \tau) \ \delta_{j+1,k}, & j \geq 0, \ k \geq 1, \nonumber \\    
L_{j,k+1} = (2 k + 1) L_{j,k} + 2 \sqrt{2 \pi} (2 j + 1)! \ 
(1 + \tau) \ \delta_{j,k}, & j \geq 0, \ k \geq 0.   
\end{eqnarray}
Using these recursion relations and noting 
\begin{equation}
L_{0,0} = - 2 \sqrt{2 \pi} (1 + \tau),
\end{equation}
we can readily find that (\ref{m1}) holds. \hfill $\square$

We remark that the two key Hermite polynomial properties (\ref{CH.4}) and
(\ref{ztimes}) allow us to verify that the first of the two formulas in
(\ref{RC}) can be rewritten to read
\begin{equation}\label{RC1b}
R_{2n+1}(z) = - (1 + \tau) e^{z^2/2(1 + \tau)}
{d \over dz} \Big ( e^{- z^2 / 2 (1 + \tau)} C_{2n}(z) \Big ).
\end{equation}
Use will be made of this form in the analysis of the correlations (\ref{rr}).

Taking the limit $\tau \to 0$ in Theorem \ref{hm1} gives the family of skew orthogonal
relevant to the real Ginibre ensemble, a result which was announced and made use of in
\cite{FN07}.

\begin{cor}\label{cw1}
The family of monic polynomials $\{p_j(z)\}_{j=0,1,\dots}$ specified by
\begin{equation}\label{C1}
p_{2n+1}(z) = z^{2n+1} - 2n z^{2n-1}, \qquad p_{2n}(z) = z^{2n}
\end{equation}
are skew orthogonal with respect to the skew inner product (\ref{fgb}) in the
case $\tau = 0$. The corresponding normalization is given by
\begin{equation}\label{C1u}
u_n := \langle p_{2n}, p_{2n+1} \rangle = (2n)! 2 \sqrt{2 \pi}.
\end{equation}
\end{cor}

\section{Asymptotic properties of the correlations}
\setcounter{equation}{0}
\subsection{Eigenvalue support}
The eigenvalue support for the real Ginibre ensemble ($\tau = 0$ case) is to leading
order a circle of radius $\sqrt{N}$.
To gain
some insight into its expected form for $0 \le \tau < 1$, consider the portion of (\ref{5.1}) which is
dependent on $\{z_j := x_j + i y_j \}_{j=1,\dots,(N-k)/2}$. For $y_j$ large this portion
is proportional to (\ref{11.ell}) with $N \mapsto (N-k)/2$. As remarked below the latter
equation, previous analysis of the one-point correlation has revealed that for the
PDF (\ref{11.ell}) the density is supported on an ellipse with semi-axes $A$
and $B$ given by (\ref{AB}). The exact results obtained above can be combined with the
analysis of \cite{DGIL94} to verify that 
this result persists in the present setting.

In \cite{DGIL94} the boundary of the support is characterized by the values of $(x,y)$ which
maximize the difference
\begin{equation}\label{2a.1}
\rho_{(1)}^{\rm c} ((x,y)) |_{N \mapsto N + 1} - \rho_{(1)}^{\rm c} ((x,y)) 
\end{equation}
for large $N$. We know from (\ref{St0}) and (\ref{St}) that
\begin{equation}\label{2a.2}
\rho_{(1)}^{\rm c} ((x,y)) = 2i e^{(y^2 - x^2)/(1 + \tau)}
{\rm erfc} \Big ( \sqrt{2 \over 1 - \tau^2} y \Big )
S_\tau^{\rm c} (\bar{z}, z).
\end{equation}
To compute the asymptotic form of (\ref{2a.1}), following the beginnings of a
strategy used to analyze the two-point correlation for (\ref{11.ell})
in \cite{FJ96}, use will be made of an integral form of
(\ref{2a.2}). 
In this regard, from a standard integral representation of the Hermite polynomials
we have
\begin{equation}\label{2a.3}
C_n(z) = {1 \over \sqrt{\pi} } 
 \int_{-\infty}^\infty  e^{ - t^2}
(z +  \sqrt{2\tau} i t)^n \, dt.
\end{equation}
It follows from this and Corollary \ref{cw1} that
\begin{equation}\label{2a.3u}
S_\tau^{\rm c}(w,z) = {1 \over \pi (1 + \tau) }
\int_{-\infty}^\infty dt_1 \, e^{- t_1^2} \int_{-\infty}^\infty dt_2 \, e^{- t_2^2} \,
S_0^{\rm c}(w + \sqrt{2 \tau} i t_1, z  +  \sqrt{2\tau} i t_2).
\end{equation}
On the other hand, substituting (\ref{C1}) in (\ref{St}) gives for $S_0^{\rm c}$ the simple
expression
\begin{equation}\label{2a.4}
S_0^{\rm c}(w,z) = {w - z \over 2 \sqrt{2 \pi} }
\sum_{j=0}^{N-2} { (wz)^j \over \Gamma(j+1) }.
\end{equation}
Substituting (\ref{2a.4}) in (\ref{2a.3u}) and making further use of (\ref{2a.3})
it follows that for large $N,\, x, \, y$
\begin{eqnarray*}
&&
\rho_{(1)}^{\rm c} ((x,y)) |_{N \mapsto N + 1} - \rho_{(1)}^{\rm c} ((x,y))
\nonumber \\
&& \qquad
\: \sim \: {\sqrt{2} \over \pi (1 + \tau) }
{1 \over (N-2)! } \exp \Big ( - {1 \over 1 - \tau^2} \Big ( (x^2 + y^2) - \tau (x^2 - y^2)
\Big ) \Big ) \Big | C_{N-2} (z) \Big |^2.
\end{eqnarray*}
This same function of $x,y$ and $N$ results from studying the difference (\ref{2a.1}) in
the case of the PDF (\ref{11.ell}). As remarked above, working in \cite{DGIL94} deduces
from this that the boundary of the support is given by an ellipse with semi-axes specified
by (\ref{AB}).

\subsection{Density of real eigenvalues}
Next asymptotic properties of the density of real eigenvalues will be considered.
According to (\ref{rr}) and (\ref{12.Sr})
\begin{equation}\label{rpra}
\rho_{(1)}^{\rm r}(x) = {e^{-x^2/2(1 + \tau)} \over 2 \sqrt{2 \pi} (1 + \tau) }
\sum_{k=0}^{N/2 - 1} {1 \over (2k)!} \Big (
\Phi_{2k}(x) R_{2k+1}(x) - \Phi_{2k+1}(x) R_{2k}(x) \Big ).
\end{equation}
The mean number of real eigenvalues is obtained by integrating $\rho_{(1)}^{\rm r}(x)$
over the real line. Making use of (\ref{rpra}), (\ref{rr1}) (with $p_k$ replaced by $R_k$), (\ref{RC1b}) 
and Theorem \ref{hm1} shows
\begin{equation}\label{rtt}
\int_{-\infty}^\infty \rho_{(1)}^{\rm r}(x) \, dx =
\sqrt{ \tau \over \pi} 
\sum_{k=0}^{N/2 - 1} {1 \over (2k)!}
\Big ( {\tau \over 2} \Big )^{2k}
\int_{-\infty}^\infty e^{- 2 \tau x^2/(1 + \tau)}
\Big ( H_{2k}(x) \Big )^2 \, dx.
\end{equation}
Further, use of a tabulated integral \cite[\S 7.373]{GR94}, and a Kummer transformation for
${}_2 F_1$, gives
\begin{eqnarray}
&&\int_{-\infty}^\infty e^{- 2 \tau x^2/ (1 + \tau) } (H_{2k}(x))^2 \, dx =
2^{2k - 1/2} \Big ( {1 + \tau \over 1 - \tau} \Big )^{1/2} \tau^{-2k - 1/2}
\Gamma(2k + 1/2)  \nonumber \\
&& \qquad \times
{}_2 F_1(1/2, 1/2; -2k + 1/2;- \tau/(1 - \tau)).
\end{eqnarray}
For large $k$ the ${}_2 F_1$ function is to leading order equal to unity.
Hence the leading order behaviour of a general term in (\ref{rtt}) is
$$
\sqrt{1 \over 2 \pi} \Big ( {1 + \tau \over 1 - \tau} \Big )^{1/2}
{1 \over (2k)^{1/2} }
$$
and so for large $N$
\begin{equation}\label{de}
\int_{-\infty}^\infty \rho_{(1)}^{\rm r}(x) \, dx \: \sim \:
\sqrt{2 N \over  \pi } \Big ( {1 + \tau \over 1 - \tau} \Big )^{1/2}.
\end{equation}

To analyze the density itself
for large $N$, we again make use of (\ref{rr1}) (with $p_k$ replaced by $R_k$),
as well as the identity
$$
R_{2k+2}(x) - (2k + 1) R_{2k}(x) = - (1 + \tau) e^{x^2/2(1+ \tau)}
{d \over dx} \Big ( e^{ - x^2/2(1 + \tau)} C_{2k + 1}(x) \Big )
$$
(cf.~(\ref{RC1b}); this can be verified using (\ref{CH.4}), (\ref{ztimes})) to rewrite
(\ref{rpra}) in the simplified form
\begin{equation}\label{rttA}
\rho_{(1)}^{\rm r}(x) =
{e^{-x^2/(1 + \tau)} \over \sqrt{2 \pi} }
\sum_{k=0}^{N-2} {1 \over k!} (C_k(x))^2 +
{ e^{- x^2/2(1 + \tau)} \over 2 \sqrt{2 \pi} (1 + \tau) }
{C_{N-1}(x) \Phi_{N-2}(x) \over (N-2)!}.
\end{equation}
Use can now be made of the classical summation formula
\begin{equation}\label{rttB}
\sum_{k=0}^\infty {t^k H_k(x) H_k(y) \over k! 2^k } =
(1 - t^2)^{-1/2} e^{- t^2(x^2 + y^2)/(1 - t^2)}
e^{2xyt/(1 - t^2)}, \qquad |t| < 1
\end{equation}
to conclude
\begin{equation}\label{dd}
\rho_{(1)}^{\rm bulk}(x) :=
\lim_{N \to \infty} \rho_{(1)}^{\rm r}(x) = {1 \over \sqrt{2 \pi (1 - \tau^2)}}.
\end{equation}
With the leading order support of the real eigenvalues the interval $[-\sqrt{N}(1+\tau),
\sqrt{N}(1+\tau)]$, to leading order the mean number of eigenvalues must be equal to
$2 \sqrt{N} (1 + \tau)$ (the length of this interval) times the density (\ref{dd}).
This reclaims (\ref{de}).

We turn our attention now to
the neighbourhood of the spectrum edge. In the case
of the real Ginibre ensemble $(\tau = 0)$ the explicit form of the density profile
about the spectrum edge at $x = \sqrt{N}$ was exhibited as \cite{FN07}
\begin{equation}\label{5.13}
\lim_{N \to \infty} \rho_{(1)}^{\rm r}(\sqrt{N} + X) \Big |_{\tau = 0}
= {1 \over \sqrt{2 \pi} } \Big (
{1 \over 2} (1 - {\rm erf} \, \sqrt{2} X ) +
{e^{- X^2} \over 2 \sqrt{2} } (1 + {\rm erf} \, X ) \Big ).
\end{equation}
For general $0 \le \tau < 1$ the density at the spectrum edge is analyzed by
setting $x = (1 + \tau) \sqrt{N} + X$ in (\ref{rttA}) then taking the limit
$N \to \infty$. As is distinct from the bulk scaling, the summation and the term
distinct from the summation both give O(1) contributions. Consider first the latter.

To calculate the explicit form of the contributions,
our main tool is the Plancherel-Rotach asymptotic formula \cite{PR34,Do07}
\begin{eqnarray}\label{PR1}
&&H_n(x) \sim (2n)^n \exp \Big ( {x^2 - x \sqrt{x^2 - 2n} - n \over 2} -
n \log (x - \sqrt{x^2 - 2n} ) \Big ) \nonumber \\
&& \qquad \times
\sqrt{{1 \over 2} \Big ( 1 + {x \over \sqrt{x^2 - 2n} } \Big ) }
\end{eqnarray}
valid for $n$ large and $x > \sqrt{2n}$.
Recalling (\ref{CH}) we require this formula with $x = \sqrt{N/2}(\sqrt{\tau} +
1/\sqrt{\tau}) + X/\sqrt{2\tau}$, $n = N - k$. A straightforward but tedious
calculation gives that for $k$ fixed
\begin{eqnarray}\label{HNk}
&& H_{N-k}(x) \: \sim \: (2 N)^{N-k} e^{-k} (1 - \tau)^{-1/2} \nonumber \\
&& \: \: \times
\exp \Big ( {N \over 2} \Big ( \tau - \log 2N - \log \tau \Big ) +
\sqrt{N} X - {X^2 \over 2 (1 - \tau) } + k + {k \over 2} \log 2N +
 {k \over 2} \log \tau \Big ).
\end{eqnarray}
Consider now the final term in (\ref{rttA}). The asymptotic form of
$C_{N-1}(x)$ can be read off (\ref{PR1}) by setting $k=1$.
To use it 
to  deduce the asymptotic form of
$\Phi_{N-2}(x)$ we require the integral evaluation \cite{GR94}
$$
\int_{-\infty}^\infty e^{-x^2} H_{2m}(xy) \, dx =
\sqrt{\pi} {(2m)! \over m!} (y^2 - 1)^m.
$$
This formula allows us write
\begin{eqnarray} 
\Phi_{N-2}(x) = \Big ( {\tau \over 2} \Big )^{N/2 - 1} \Big (
\sqrt{2 \pi (1 + \tau)} {(N-2)! \over (N/2 - 1)!} \tau^{1 - N/2} \nonumber \\
\qquad - 2 \int_x^\infty e^{-t^2/2(1+\tau)} H_{N-2} \Big ( {t \over \sqrt{2\tau} }
\Big ) \, dt \Big ).
\end{eqnarray}
It is in this form that we substitute (\ref{HNk}) with $k=2$. 
Combining results and 
making use too of
Stirling's formula allows us to compute the sought limiting form, 
\begin{eqnarray}\label{Ad1}
&& \lim_{N \to \infty} \Big ( 
{ e^{- x^2/2(1 + \tau)} \over 2 \sqrt{2 \pi} (1 + \tau) }
{C_{N-1}(x) \Phi_{N-2}(x) \over (N-2)!} \Big |_{x = (1 + \tau)\sqrt{N} + X}
\Big ) \nonumber \\
&& \qquad =
{1 \over (1 - \tau^2)^{1/2} }
{e ^{- X^2/(1 - \tau^2)} \over 4 \sqrt{\pi} }
\Big ( 1 + {\rm erf} (X/ \sqrt{1 - \tau^2}) \Big ).
\end{eqnarray}

To analyze the sum in (\ref{rttA}) the asymptotic expansion (\ref{HNk}) must be extended
to include terms O$(k/\sqrt{N})$. One finds these terms to be the multiplicative factor
$$
\exp \Big ( - {k^2 \tau \over 2 N (1 - \tau)} - {k X \over \sqrt{N} (1 - \tau) } \Big ).
$$
Noting too from Stirling's formula that
$$
{1 \over (N - k)! } \: \sim \:
(2 \pi N)^{1/2} \exp \Big ( N \log N - N - k \log N + {k^2 \over 2 N} \Big ),
$$
after rearranging the order of summation so that $k \mapsto N - k$, $(k=2,\dots,N)$ 
and recognizing that a Riemann sum approximation to a definite integral results
we find
\begin{eqnarray}\label{Ad2}
&& \lim_{N \to \infty}
{ e^{- x^2/(1 + \tau)} \over \sqrt{2 \pi} }
\sum_{k=0}^{N-2} {1 \over k!} (C_k(x))^2 \Big |_{x = (1 + \tau) \sqrt{N} + X} 
\nonumber \\
&& \qquad
=
{1 \over (1 - \tau^2)^{1/2} } {1 \over 2 \sqrt{2 \pi} }
\Big (1 - {\rm erf } ( \sqrt{2} X/ (1 - \tau^2)^{1/2} ) \Big ).
\end{eqnarray}
Now
adding together (\ref{Ad1}) and (\ref{Ad2}) gives for the edge density
\begin{eqnarray}\label{Ad3}
&& \rho_{(1)}^{\rm edge}(X) := \lim_{N \to \infty}
\rho_{(1)}^{\rm r} ((1 + \tau) \sqrt{N} + X) =
{1 \over \sqrt{2 \pi (1 - \tau^2)} }
\nonumber \\
&& \qquad
\times  \Big (
{1 \over 2} \Big (1 - {\rm erf} (\sqrt{2} X/(1 - \tau^2)^{1/2}) \Big )
+ { e^{- X^2/(1 - \tau^2) } \over 2 \sqrt{2} }
\Big ( 1 + {\rm erf} (X/(1 - \tau^2)^{1/2}) \Big )\Big ).
\end{eqnarray}
Note that this agrees with (\ref{5.13}) in the case $\tau = 0$.

We observe from (\ref{Ad3}) that
$\rho_{(1)}^{\rm edge}(X) dX$ for general $0 \le \tau < 1$ is obtained from
the case $\tau = 0$ by the simple scaling $X \mapsto X/\sqrt{1 - \tau^2}$. Note that
this same rule is valid for the bulk density of real eigenvalues (\ref{dd}).
Indeed it is reasonable to expect that all local correlations are only altered by this
change of scale, as the point process for general $0 \le \tau < 1$ is locally
identical to that for the point process in the case $\tau = 0$, except that 
the two-dimensional bulk density
is scaled by a factor of $1/(1 - \tau^2)$. We will now proceed to exhibit this fact
for the general real-real and complex-complex correlations in the bulk.

\subsection{$k$-point correlations in the bulk}
Consider first the complex-complex case. 
Setting
$$
\hat{S}_\tau^{\rm c}(w,z) :=
e^{ - (z^2 + w^2)/2(1 + \tau)} S_\tau^{\rm c}(w,z),
$$
we see that (\ref{St0}) can be rewritten
\begin{eqnarray}\label{Swz}
&& \rho_{(n)}^{\rm c}((x_1,y_1),\dots,(x_n,y_n)) = \nonumber \\
&& \qquad
\prod_{l=1}^n \Big ( 2 i {\rm erfc} \Big ( \sqrt{2 \over 1 - \tau^2}  y_l
\Big ) 
{\rm Pf} \left [ \begin{array}{cc} \hat{S}_\tau^{\rm c}(\bar{z}_j, \bar{z}_k) &
 \hat{S}_\tau^{\rm c}(\bar{z}_j, {z}_k) \\
\hat{S}_\tau^{\rm c}({z}_j, \bar{z}_k) & \hat{S}_\tau^{\rm c}({z}_j, {z}_k)
\end{array} \right ].
\end{eqnarray}

Now, it is immediate from (\ref{2a.4}) that
$$
\lim_{N \to \infty} S_0^{\rm c}(w,z) = {w - z \over 2 \sqrt{2 \pi} } e^{wz}.
$$
Substituting this in (\ref{2a.3}) and computing the resulting Gaussian integrals gives
\begin{equation}\label{5.5.1}
\lim_{N \to \infty} S_\tau^{\rm c}(w,z) =
{ (w - z) \over 2 \sqrt{2 \pi} (1 - \tau^2) }
\exp \Big ( - {\tau \over 2 (1 - \tau^2) } (z^2 + w^2) +
{1 \over 1 - \tau^2} zw \Big ).
\end{equation}
Consequently
\begin{equation}\label{5.22a}
\lim_{N \to \infty} \hat{S}_\tau^{\rm c}(w,z) =
{(w - z) \over 2 \sqrt{2 \pi} (1 - \tau^2) }
\exp \Big ( - {(z - w)^2 \over 2 (1 - \tau^2) } \Big ).
\end{equation}
Substituting in (\ref{Swz}) gives the bulk complex-complex correlations. The
feature that the bulk limiting value of
\begin{equation}\label{5.5.3}
\rho_{(n)}^{\rm c}((x_1,y_1),\dots,(x_n,y_n)) \, \prod_{l=1}^n dx_l  dy_l 
\end{equation}
for general $0 \le \tau < 1$ is gotten from the $\tau = 0$ case by the replacements
$$
(x_l, y_l) \mapsto (x_l/\sqrt{1 - \tau^2}, y_l/\sqrt{1 - \tau^2})
$$ 
is evident. 

It remains to consider the real-real case. Proceeding as in the derivation of
(\ref{rttA}) shows that (\ref{12.Sr}) can be rewritten
\begin{eqnarray}\label{wev}
S^{\rm r}(x,y) = { e^{- (x^2 + y^2)/2 (1 + \tau)} \over \sqrt{2 \pi} }
\sum_{k=0}^{N-2} {1 \over k!} C_k(x) C_k(y) +
{ e^{ - y^2 / 2 (1 + \tau) } \over 2 \sqrt{2 \pi} (1 + \tau) }
{C_{N-1}(y) \Phi_{N-2}(x) \over (N-2)!}
\end{eqnarray}
As for the derivation of (\ref{dd}), we use (\ref{rttB}) to both deduce that the
final term vanishes as $N \to \infty$ (a consequence of the convergence of the sum)
and to give a closed form evaluation of the summation. It follows that
\begin{equation}\label{5.25}
 \lim_{N \to \infty} S^{\rm r}(x,y)  =
{1 \over \sqrt{2 \pi (1 - \tau^2) } } e^{- (x - y)^2/2(1 - \tau^2) }.
\end{equation}
In view of the formulas (\ref{12.Sr1}) for the remaining quantities in 
(\ref{rr}), as for (\ref{5.5.3}) we have that the limiting bulk value of
$$
\rho_{(n)}^{\rm r}(x_1,\dots,x_n) \, dx_1 \cdots dx_n 
$$
for general $0 \le \tau < 1$ is gotten from the $\tau = 0$ case by the replacements
$x_j \mapsto x_j / \sqrt{1 - \tau^2}$.

\subsection{The weakly non-symmetric limit}
In relation to the ensemble interpolating between the complex Ginibre
ensemble and the GUE, it was exhibited in \cite{FKS97} that well defined
correlations result by setting $\tau = 1 - \alpha^2/ N$, then taking
$N \to \infty$. Similarly, scaled correlations were computed in this
limit for the ensemble interpolating between the real quaternion Ginibre
ensemble and the GSE. Here the scaled correlations of the real Ginibre/
GOE interpolating ensemble will be calculated.
Note that with this scaling (\ref{AB}) gives that the eigenvalue support
collapses onto the interval $[ - 2 \sqrt{N}, 2 \sqrt{N}]$ of the real axis.
The mean spacing between eigenvalues is then O$(1/\sqrt{N})$, suggesting that 
we should also multiply coordinates by $\pi /\sqrt{N}$ (the 
proportionality $\pi$ is chosen for convenience; then a unit real density
results) before taking $N \to \infty$.

Now, using the asymptotic expansion
$$
{\Gamma (n/2 + 1) \over \Gamma (n + 1)} e^{- x^2} H_n(x) =
\cos \Big ( \sqrt{2 n + 1} x - n \pi /2 \Big ) + {\rm O}(n^{-1/2}),
$$
we deduce from (\ref{wev}) that
$$
{\pi \over \sqrt{N} } S_\tau^{\rm r} \Big ( {\pi x \over \sqrt{N}},
{\pi y \over \sqrt{N} } \Big ) \Big |_{\tau = 1 - \alpha^2/2N}
\sim {1 \over 2} \sqrt{\pi \over 2 N}
\sum_{k=0}^{N-2} { e^{- k \alpha^2/N} k! \over 2^k ((k/2)!)^2 }
\cos \Big ( \pi \sqrt{k \over N} (x - y) \Big ).
$$
Making use of Stirling's formula, a Riemann sum approximation to a definite integral
is obtained, and we compute
\begin{equation}\label{aca}
\lim_{N \to \infty} {\pi \over \sqrt{N} }
S_\tau^{\rm r} \Big ( {\pi x \over \sqrt{N} }, {\pi y \over \sqrt{N} }
\Big ) \Big |_{\tau = 1 - \alpha^2/N} =
\int_0^1 e^{- \alpha^2 u^2} \cos \pi u (x - y) \, du.
\end{equation}
Note that in contrast to the correlations implied by
(\ref{5.25}), we see from (\ref{aca}) that the correlations in the present
setting of the weakly non-symmetric limit exhibit an algebraic decay.
For the limiting form of the complex-complex correlations, we first note from 
(\ref{2a.3})--(\ref{2a.4}) that
$$
S_\tau^{\rm c}(w,z) = {1 \over 2 (1 + \tau) \sqrt{2 \pi} }
\sum_{j=0}^{N-2} {C_{j+1}(w) C_j(z) - C_j(w) C_{j+1}(z) \over \Gamma(j+1) }.
$$
Proceeding now as for the working which lead to (\ref{aca}) shows
\begin{equation}\label{5.27}
\lim_{N \to \infty} \Big ({\pi \over \sqrt{N} } \Big )^2
S_\tau^{\rm c} \Big ( {\pi w \over \sqrt{N} }, {\pi z \over \sqrt{N} }
\Big ) \Big |_{\tau = 1 - \alpha^2/N} =
{\pi \over 2} \int_0^1 u e^{- \alpha^2 u^2} \sin \pi u (w - z) \, du,
\end{equation}
where on the LHS we have used the fact that
the complex-complex correlations must be scaled by $(\pi/\sqrt{N})^2$
for each independent two-dimensional coordinate $(x,y)$ to account for the measure
in (\ref{5.5.3}). Substituting this in (\ref{St0}) and noting too that
$$
e^{(y^2_l - x^2_l)/(1 + \tau)} {\rm erfc} \Big (
\sqrt{2 \over 1 - \tau^2} y_l \Big ) \to
{\rm erfc} \Big ( { \pi y_l \over \alpha} \Big )
$$
gives the explicit weakly non-symmetric limiting form of $\rho^{\rm c}_{(n)}$.

\section*{Acknowledgements}
The work of PJF was supported by the Australian Research Council. We thank Alexei
Borodin for providing us with a copy of \cite{BS08} prior to posting on the arXiv.


\providecommand{\bysame}{\leavevmode\hbox to3em{\hrulefill}\thinspace}
\providecommand{\MR}{\relax\ifhmode\unskip\space\fi MR }
\providecommand{\MRhref}[2]{%
  \href{http://www.ams.org/mathscinet-getitem?mr=#1}{#2}
}
\providecommand{\href}[2]{#2}

\end{document}